# Organic versus Hybrid Coacervate Complexes :
# Co-Assembly and Adsorption Properties

Ling Qi[1], Jean-Paul Chapel[1], Jean-Christophe Castaing[1], Jérôme Fresnais[2] and Jean-François Berret[2@]

*(1) : Complex Fluid Laboratory, UMR CNRS/Rhodia 166, Rhodia North-America, R&D Headquarters CRTB,*
*350 George Patterson Blvd., Bristol, PA 19007 USA.*
*(2) : Matière et Systèmes Complexes, UMR 7057 CNRS Université Denis Diderot Paris-VII, Bâtiment Condorcet,*
*10 rue Alice Domon et Léonie Duquet, 75205 Paris, France*

We report the co-assembly and adsorption properties of coacervate complexes made from polyelectrolyte-neutral block copolymers and oppositely charged nanocolloids. The nanocolloids put under scrutiny were ionic surfactant micelles and highly charged 7 nm cerium oxide ($CeO_2$) nanoparticles. Static and dynamic light scattering was used to investigate the microstructure and stability of the organic and hybrid complexes. For five different systems of nanocolloids and polymers, we first demonstrated that the electrostatic complexation resulted in the formation of stable core-shell aggregates in the 100 nm range. The microstructure of the $CeO_2$-based complexes was resolved using cryogenic transmission electronic microscopy (Cryo-TEM), and it revealed that the cores were clusters made from densely packed nanoparticles, presumably through complexation of the polyelectrolyte blocks by the surface charges. The cluster stability was monitored by systematic light scattering measurements. In the concentration range of interest, $c = 10^{-4} - 1$ wt. %, the surfactant-based complexes were shown to exhibit a critical association concentration (cac) whereas the nanoparticle-polymer hybrids did not. The adsorption properties of the same complexes were investigated above the cac by stagnation point adsorption reflectometry. The adsorbed amount was measured as a function of time for polymers and complexes using anionically charged silica and hydrophobic poly(styrene) substrates. It was found that all complexes adsorbed readily on both types of substrates up to a level of $1 - 2$ mg $m^{-2}$ at stationary state. Upon rinsing however, the adsorbed layer was removed for the surfactant-based systems, but not for the cerium oxide clusters. As for the solution properties, these finding were interpreted in terms of a critical association concentrations which are very different for organic and hybrid complexes. Combining the efficient adsorption and strong stability of the $CeO_2$–based core-shell hybrids on various substrates, it is finally suggested that these systems could be used appropriately for coating and anti-biofouling applications.

## I – Introduction

Development of functional molecular architectures is one of the final goals of modern chemistry, biology and physics. For the design of artificially intelligent devices, fabrication and control of materials at nanometer scale with chemical and physical attributes has been attracting much attention in the last decade. In particular, the complexation of polymers and nanoparticles is opening pathways for engineering novel hybrid structures combining the advantageous properties of both the organic and inorganic moieties [1-3].

The first challenge of the present paper was the design of nano-objects with new functionalities and resulting from the association between nanocolloids and polymers. The nanocolloids put under scrutiny were of two kinds: surfactant micelles and inorganic cerium oxide nanoparticles. As for polymers, polyelectrolyte-neutral block copolymers also called double-hydrophilic copolymers were employed [4]. The association was driven by the electrostatic interactions between the oppositely charged polyelectrolytes and particles [5-13]. In the context of electrostatic complexation, it has been shown that polyelectrolyte-neutral copolymers exhibited interesting clustering and coating properties. These hydrosoluble macromolecules were found to co-assemble with oppositely charged systems, e.g. with surfactants [14-16], polymers [17-22] and proteins [23-27], yielding "supermicellar" aggregates with core-shell structures. As a result, the core of the aggregates was described as a dense coacervate microphase comprising the oppositely charged species. Made from neutral blocks, the corona was identified as surrounding the cores and related to the stability of the whole colloid. In the present paper, the same type of formulations were conducted, using on one hand ionic surfactants such as dodecyl trimethylammonium bromide (DTAB) and sodium dodecyl sulfate (SDS) and on the other cerium oxide nanoparticles

($CeO_2$, nanoceria). Nanoceria were considered because of their potential applications in catalysis [28-30] polishing and coating technologies [31,32] as well as in biology [33,34].

The second goal of this survey was to search for synergistic complexation effects in the context of coating and anti-biofouling applications. Cohen Stuart and coworkers have shown that electrostatic complexes made from oppositely charged polyelectrolytes could be efficiently adsorbed on charged hydrophilic and on hydrophobic surfaces, such as silica and poly(styrene) respectively [20,21,35]. Stemming from systems without hydrophobic moieties, the affinity to poly(styrene) substrates was a rather surprising result. It was attributed to an entropy effect and to a minimization of the contact area between poly(styrene) and the solvent. It was concluded that on a microscopic level, the cores of the aggregates adsorbed on the surface whereas the shell formed a brush on the top of it. Cohen Stuart and coworkers have also demonstrated the stability of the deposited layer upon rinsing, and its remarkable anti-fouling properties with addition of proteins [20,21,35]. Adsorptions on the same substrates of electrostatic coacervate phases were also obtained with surfactants and polyelectrolytes were also reported recently [36-38]. These studies have revealed interesting co-adsorption features and non equilibrium behaviors at the liquid-solid interfaces. The approach followed here has consisted to extend the adsorption measurements to new types of electrostatic complexes, namely to complexes built from surfactant micelles and from cerium oxide nanoparticles. Doing so, we anticipated the possibility to combine the efficient adsorption already observed and the strong anti-UV filter properties of nanoceria. Here, we show the existence of a correlation between the bulk and adsorption properties of new polymer-coated nanoclusters [39]. We have found that the affinities for the poly(styrene) and silica surfaces were stronger for hybrids than those from surfactant-polymer complexes. Their superior stability and remanence, even after rinsing were related to the fact that nanoparticle-polymer complexes display extremely low critical association concentration.

| specimen | $M_w$ (Kg mol$^{-1}$) | dn/dc[a] (cm$^3$ g$^{-1}$) | $D_H$[b] (nm) | Refs. |
|---|---|---|---|---|
| PANa$_{5K}$-*b*-PAM$_{30K}$ | 34.7 | 0.157 | 11 | 59 |
| PSS$_{7K}$-*b*-PAM$_{30K}$ | 69.0 | 0.158 | 19 | 59 |
| PTEA$_{11K}$-*b*-PAM$_{30K}$ | 35.4 | 0.153 | 11 | 59 |
| DTAB micelles | 16.4 | 0.231 | 4 | 42 |
| SDS micelles | 14.4 | 0.152 | 4 | 42 |
| CeO$_2$ (acidic pH) | 330 | 0.195 | 9.8 | (this work) |
| PAA$_{2K}$ coated CeO$_2$ | 430 | 0.184 | 13 | (this work) |

**Table I** : *Weight-average molecular weight MW, refractive index increment dn/dc and hydrodynamic diameter DH of the block copolymers, surfactant micelles and nanoparticles studied in this work. (a) : measured at $\lambda$ = 632.8 nm. (b) : hydrodynamic diameter determined from cumulant analysis (second cumulant coefficient) and extrapolated at c $\rightarrow$ 0.*

## II – Experimental

### II.1 – Chemicals, Characterization and Sample Preparation

*II.1.1 – polymers, surfactant and nanoparticles*

Polymers : For electrostatic complexation, we have used three different block copolymers, poly(sodium acrylate)-*b*-poly(acrylamide), poly(styrene sodium sulfonate)-*b*-poly(acrylamide) and poly(trimethylammonium ethylacry-late methylsulfate-*b*-poly(acrylamide), abbreviated in the following as PANa-*b*-PAM, PSS-*b*-PAM and PTEA-*b*-PAM respectively. The diblock copolymers were synthesized by controlled radical polymerization according to MADIX technology [40,41] and the chemical formulae of the monomers are given in previous reports [16]. In the present study, we focus on three different molecular weights, PANa$_{5K}$-*b*-PAM$_{30K}$, PSS$_{7K}$-*b*-PAM$_{30K}$ and PTEA$_{11K}$-*b*-PAM$_{30K}$. The values in indices are the molecular weights targeted by the synthesis, and except for PSS$_{7K}$-*b*-PAM$_{30K}$ these numbers are in good agreement with those determined experimentally. In aqueous solutions and neutral pH conditions, the chains are well dispersed and in the state of unimers. Light scattering has been performed on dilute solutions, and the weight-averaged molecular weights and the hydrodynamic diameter determined (Table I). The polydispersity index was determined by size exclusion chromatography at 1.6.

In addition to the block copolymers, we have also used homopolymers (Sigma Aldrich) such as poly(acrylic acid) with a weight-averaged molecular weight of 2000 g mol$^{-1}$ and poly(acrylamide) with molecular weight 10000 g mol$^{-1}$. The polymers were used as received. When required, the solution pHs were adjusted by titration using reagent-grade nitric acid (HNO$_3$), ammonium hydroxide (NH$_4$OH) or sodium hydroxide (NaOH).

Surfactant : Dodecyltrimethylammonium bromide (DTAB) and sodium dodecylsulfate (SDS) were purchased from Sigma and used without further purification. Both surfactants possess a C12 aliphatic chain and are characterized by critical micellar concentrations (cmc) in water 0.46 wt. % (15 mMol) for DTAB and 0.23 wt. % (8 mMol) for SDS. The aggregation numbers for DTAB and SDS micelles were evaluated by small-angle neutron scattering at $N_{Agg}$ = 53 and $N_{Agg}$ = 50, respectively.42 These numbers were found to be in good agreement with literature data on the same systems [43-45].

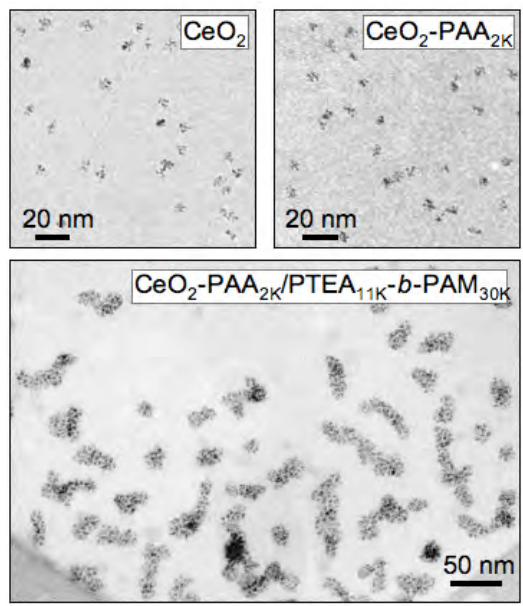

**Figure 1** : *Cryogenic transmission electron microscopy (cryo-TEM) images of bare and coated cerium oxide nanoparticles. Upper left panel : bare CeO2 in acidic conditions (pH 1.4); Upper right panel : CeO2 coated with poly(acrylic acid) with molecular weight 2000 g mol-1 (pH 8). Lower panel : CeO2-PAA2K complexed with the oppositely charged diblock copolymers PTEA11K-b-PAM30K.*

Nanoparticle : The nanoparticle system investigated in this work was a dispersion of cerium oxide nanocrystals, or nanoceria. The synthetic procedure involved thermohydrolysis of an acidic solution of cerium-IV nitrate salt at high temperature, which resulted in the homogeneous precipitation of a cerium oxide pulp [10]. The size of the particles was controlled by addition of hydroxide ions during the thermohydrolysis. High resolution transmission electron microscopy have shown that the nanoceria (bulk mass density $\rho$ = 7.1 g cm$^{-3}$) consisted of isotropic agglomerates of 2 - 5 crystallites with typical size 2 nm and faceted morphologies. Wide-angle x-ray scattering confirmed the crystalline fluorite structure of the nanocrystallites [10]. Cryo-TEM images of single nanoparticles are displayed in Fig. 1a (upper left panel). An image analysis performed on 350 particles has allowed us to derive the size distribution function, which was found to be well-accounted for by a log-normal function [13].

$$p(D, D_0, s) = \frac{1}{\sqrt{2\pi} \beta(s) D} \exp\left(-\frac{\ln^2(D/D_0)}{2\beta(s)^2}\right) \quad (1)$$

where $D_0$ (= 6.9 $\pm$ 0.3 nm) is the median diameter and where $\beta(s)$ is related to the polydispersity index s by the relationship $\beta(s) = \sqrt{\ln(1+s^2)}$. The polydispersity index (s = 0.15 $\pm$ 0.03) is defined as the ratio between the standard deviation and the average diameter [46]. From this distribution, the radius of gyration and the hydrodynamic diameter were calculated to be $R_G = 3.5$ nm and $D_H = 8.6$ nm, in good agreement with the direct x-ray and light scattering determinations (which are at 3.5 nm and 9.8 nm, respectively [10,13]).

*II.1.2 – Sample Preparation*

In order to adsorb ion-containing polymers onto nanoceria, we have followed the precipitation-redispersion pathway recently described in the literature.10 The precipitation of the cationic $CeO_2$ dispersion by $PAA_{2K}$ was first performed in acidic and dilute conditions. As the pH of the solution was then increased by addition of ammonium hydroxide, the precipitate redispersed spontaneously, yielding a clear solution. As shown in Table I, the hydrodynamic size found in $CeO_2$-$PAA_{2K}$ dispersions was $D_H = 13$ nm, i.e. 3.2 nm above that of the bare particles. This increase was interpreted as arising from a charged $PAA_{2K}$ brush surrounding the particles. Quantitative determinations using small-angle scattering experiments and chemical analysis have disclosed a brush comprising ~ 50 $PAA_{2K}$ chains adsorbed at the interfaces. A cryo-TEM image of the $PAA_{2K}$-coated nanoceria is displayed in Fig. 1b (upper right panel). This technique revealed isolated particles with size and size distribution identical to those obtained for the bare acidic particles. Due to the low electronic contrast between polymer and water, the presence of organic $PAA_{2K}$ brushes around the nanoceria could not be directly inferred by cryo-TEM.

Mixed solutions of nanocolloids and copolymers were prepared by simple mixing of dilute solutions prepared at the same concentration c (c ~ 1 wt. %) and same pH. As mentioned already in the introduction, the term "nanocolloid" here applies to surfactant micelles as well as to nanoceria particles. The relative amount of each component was monitored by the charge ratio Z :

$$Z = \frac{Q[Nano]}{n_{PE}[Pol]} \qquad (2)$$

where [Nano] and [Pol] were the molar nanocolloid and polymer concentrations, Q the average structural charge borne by the colloid and $n_{PE}$ the degree of polymerization of the polyelectrolyte block. For DTAB and SDS micelles, the structural charge was assimilated to the number of surfactants per micelle (i.e. the aggregation number), yielding $Q_{DTAB}$ = +53e and $Q_{SDS}$ = -50e. e denotes here the elementary charge. For the bare and coated cerium particles, we have estimated the structural charge at $Q_{CeO2}$ = +300e (at acidic pH) [28,47] and $Q_{CeO2-PAA2K}$ = -700e (at neutral pH) [10]. This latter estimate was made assuming that half of the carboxylate groups of $PAA_{2K}$ are adsorbed on the surface of the particles, and the remaining are dangling in the solvent, forming then the charged corona. For the present study, organic and hybrid complexes were prepared at Z = 1, i.e. for solutions characterized by the same number densities of positive and negative chargeable ions [35,48].

**II.2 – cryogenic transmission electron microscopy**

Cryo-transmission electron microscopy (cryo-TEM) was performed on particle and polymer-nanoparticle solutions prepared at concentration c = 0.2 wt. % (Z = 1). For the experiments, a drop of the solution was put on a TEM-grid covered by a 100 nm-thick polymer perforated membrane. The drop was blotted with filter paper and the grid was quenched rapidly in liquid ethane in order to avoid the crystallization of the aqueous phase. The membrane was then transferred into the vacuum column of a TEM-microscope (JEOL 1200 EX operating at 120 kV) where it was maintained at liquid nitrogen temperature. The magnification for the cryo-TEM experiments was selected at 40 000×.

**II.3 – Static and dynamic light scattering**

Static and dynamic light scattering were performed on a Brookhaven spectrometer (BI-9000AT autocorrelator, $\lambda$ = 488 nm) for measurements of the Rayleigh ratio and of the collective diffusion constant D(c). The Rayleigh ratio was obtained from the scattered intensity I(q,c) measured at the wave-vector q according to [49] :

$$\mathcal{R}(q,c) = \mathcal{R}_{std} \frac{I(q,c) - I_S}{I_{Tol}} \left( \frac{n}{n_{Tol}} \right)^2 \qquad (3)$$

In Eq. 3, $\mathcal{R}(q,c)$ and $n_{Tol}$ are the standard Rayleigh ratio and refractive index of toluene, $I_S$ and $I_{Tol}$ the intensities measured for the solvent and for the toluene in the same scattering configuration and $q = \frac{4\pi n}{\lambda} \sin(\theta/2)$ (with n the refractive index of the solution and the scattering angle). Light scattering was used to determine the apparent molecular weight $M_{w,app}$ and radius of the gyration $R_G$ of the macromolecules and colloids investigated here. In the regime of weak colloidal interactions, the Rayleigh ratio was found to follow a wave-vector and concentration dependence which is highlighted by the Zimm representation [49] :

$$\frac{Kc}{\mathcal{R}(q,c)} = \frac{1}{M_{w,app}} \left( 1 + \frac{q^2 R_G^2}{3} \right) + 2A_2 c \qquad (4)$$

In Eq. 4, $K = 4\pi^2 n^2 (dn/dc)^2/N_A\lambda^4$ is the scattering contrast coefficient ($N_A$ is the Avogadro number) and $A_2$ is the second virial coefficient. The refractive index increments dn/dc of the different solutions were measured using an Agilent 1100 RID refractive index detector (deflection method - Agilent Technologies, Santa Clara, CA) in the range c = $10^{-3}$ – 0.1 wt. % and with a Bellingham + Stanley RFM 830 refractometer for c = 0.1 – 1 wt. %. The values of the refractive index increments for the polymers, surfactant and nanoparticles solutions are shown in Table I. For the polymers and the nanoparticles in the dilute concentration range (c < 0.2 wt. %), $qR_G$ << 1 and Eq. 2 reduces to . This latter equation emphasizes the fact that for small sizes, the Rayleigh ratio does not depend on the wave-vector in the window $6 \times 10^{-4} - 4 \times 10^{-3}$ Å$^{-1}$ characteristic for light scattering. In quasi-elastic light scattering, the collective diffusion coefficient D(c) was measured in the range c = 0.01 wt. % – 1 wt. %. From the value of D(c) extrapolated at c = 0 (noted D0), the hydrodynamic radius of the colloids was calculated according to the Stokes-Einstein relation, $D_H = k_B T/3\pi\eta_S D$, where $k_B$ is the Boltzmann constant, T the temperature (T = 298 K) and $\eta_S$ ($\eta_S = 0.89 \times 10^{-3}$ Pa s) the solvent viscosity. The autocorrelation functions of the scattered light were interpreted using both the method of cumulants and the CONTIN fitting procedure provided by the instrument software.

**II.4 – Adsorption on surfaces monitored by reflectometry**

*Optical reflectometry* : The amount of adsorbed polymers and coacervates deposited onto silica or poly(styrene) (PS) surfaces was monitored using Stagnation Point Adsorption Reflectometry (SPAR) [50,51]. Fixed angle reflectometry measured the reflectance at the Brewster angle on the flat substrate. A linearly polarized light beam was reflected by the surface and subsequently splitted into a parallel and a perpendicular component using a polarizing beam splitter. As material adsorbed at the substrate-solution interface, the ratio S between the parallel and perpendicular components of the reflected light varied [51]. The system was analyzed in terms of Fresnel reflectivities for a multi-layer system (substrate, coating, adsorbed layer and solvent), where each layer was described by a complex refractive index and a layer thickness. According to this model, the sensitivity factor ($A_S$), which is the relative change in the output signal S per unit surface was found to be proportional to dn/dc. It also depended on the angle of incidence, the wavelength and the nature of the collecting surface. The change in S was then related to the adsorbed amount through [21,51] :

$$\Gamma(t) = \frac{1}{A_s} \frac{S(t) - S_0}{S_0} \qquad (5)$$

where $S_0$ is the signal from the bare surface prior to adsorption. Adsorption measurements were obtained as a function of the time with a 100 ms-resolution up to a stationary state $\Gamma_{ST}$.

The parameter $A_S$ in Eq. 5 was found to be very weakly dependent upon the amount of material adsorbed. In practice, it was regarded as a constant. Furthermore, a good accuracy and repeatability were obtained when $A_S$ is larger than 0.005 $m^2$ $mg^{-1}$. The reflectometer set-up was equipped with a stagnation-point flow cell in which hydrodynamics conditions are well-defined. At the stagnation point the hydrodynamic flow was zero leading to diffusion-limited exchanges between the injected solution and the collecting surface. A complete description of the stagnation point adsorption reflectometry device can be found in several references, e.g. in Refs. [21,38,51-54].

*Model substrates* : Hydrophilic silica substrates were modeled using smooth silicon wafers covered with a layer of ~ 100 nm $SiO_2$ (from Silicon Inc. Bose, Idaho) in order to maximize the reflectometer signal. The surface was cleaned under UV radiation for 15 mn, and then rinsed with de-ionized (DI) water just before being introduced in the reflectometer cell. At this stage, the surface was highly hydrophilic showing a zero water contact angle. Hydrophobic poly(styrene) substrate was modeled by a poly(styrene) thin layer of ~ 100 nm deposited on top of an HMDS (hexamethyldisilane) functionalized silicon wafer by spin-coating a toluene solution (2.5 wt. %) at 5000 rpm. The silane treatment was meant to avoid a dewetting of the PS layer with time or during the annealing step (2 hours at 100° C). The final PS layer thickness was around 100 nm to insure a good sensitivity and presented water contact angles around 88° typical for PS coated material. The exact thickness and refractive index of the substrates (silica or poly(styrene)) were measured by spectroscopic ellipsometry (ESG4 from SOPRA, France) prior to the reflectometry experiment in order to determine the sensitivity factor AS and the adsorbed amount $\Gamma$ from the optical model described further up. The value for the refractive index increments are given in Table I. The sensitivity factor AS were found to range between 0.02 and 0.035 $m^2$ $mg^{-1}$ for silica surfaces and between 0.017 and 0.025 $m^2$ $mg^{-1}$ for PS surfaces.

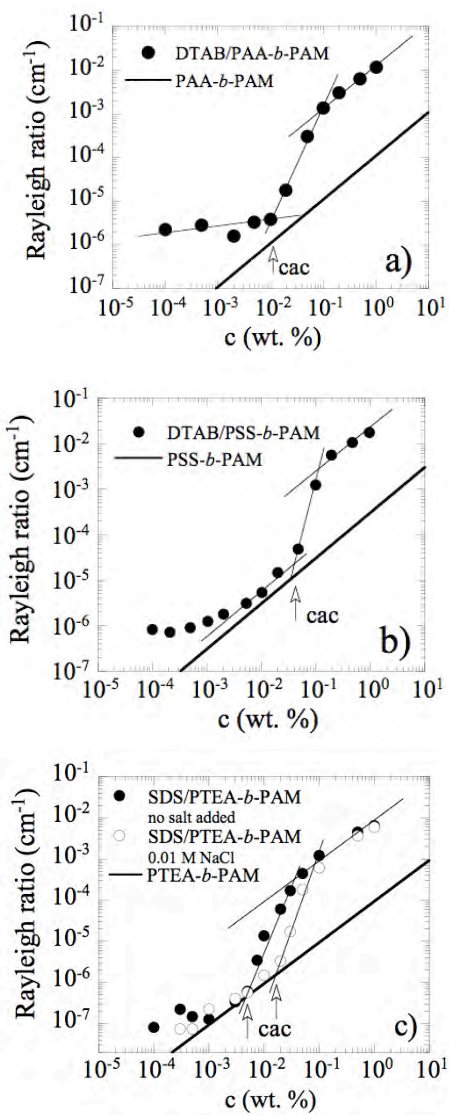

*Figure 2 : Concentration dependences of the Rayleigh ratios obtained by light scattering (in the 90°-configuration) for a) DTAB/PANa5K-b-PAM30K, b) DTAB/PSS7K-b-PAM30K and c) SDS/PTEA11K-b-PAM30K. For this latter system, dilution lines with and without added salt (0.01 M NaCl) were studied. The critical association concentrations (cac) are indicated by an arrow and their values are given in Table II. The thick straight lines depict the intensities calculated assuming that surfactants and polymers were not associated.*

## III – Results and Discussion

### III.1 – Bulk Properties of Complex Coacervates

*III.1.1 - Critical Aggregation Concentration of Organic Complexes*

Figs. 2 display the concentration dependences of the Rayleigh ratio obtained by light scattering (90°-configuration) for DTAB/PANa$_{5K}$-b-PAM$_{30K}$, DTAB/PSS$_{7K}$-b-PAM$_{30K}$ and SDS/PTEA$_{11K}$-b-PAM$_{30K}$. The surfactant-polymer complexes in Figs. 2 were prepared at total weight concentration c = 1 wt. % and charge ratio Z = 1. They were further divided by addition of water or brine. For SDS/PTEA$_{11K}$-*b*-PAM$_{30K}$ (Fig. 2c), two dilution lines were investigated, one without added salt (close symbols) and one with 0.01 M NaCl (open symbols). With decreasing concentration, the Rayleigh ratios exhibited first a linear dependence, in agreement with the dilution law of colloidal systems (Eq. 4). Below a concentration c = $10^{-2}$ – $10^{-1}$ wt. %, the scattering intensity decreased rapidly before attaining a second linear dependence. By analogy with the micellization of surfactants,[55] this second change of slope was interpreted in terms of a critical association concentration (cac) [56]. For the three systems of Figs. 2, one obtained $c_{cac}$ = $1.1×10^{-2}$, $4.2×10^{-2}$ and $5.0×10^{-3}$ wt. % respectively. Binding isotherms using electrodes selective for DTAB molecules corroborated the preceding data. Performed at low concentration on PAA-*b*-PAM copolymers, potentiometric studies revealed a cac of (3 ± 1)×$10^{-3}$ wt. %, in relative agreement with the data of Fig. 2a. At still lower concentration, i.e. below $10^{-3}$ wt. %, the intensity reached a concentration independent plateau around $\mathcal{R}$ ~ $10^{-7}$ cm$^{-1}$. This latter value represented the minimum Rayleigh ratio that could be detected with this technique. These different regimes have been highlighted by thin straight lines in Fig. 2. Note that in agreement with the thermodynamic complexation model developed by Konop and Colby [57] the cac for SDS/PTEA$_{11K}$-*b*-PAM$_{30K}$ in 0.01 M brine is higher than in pure water.

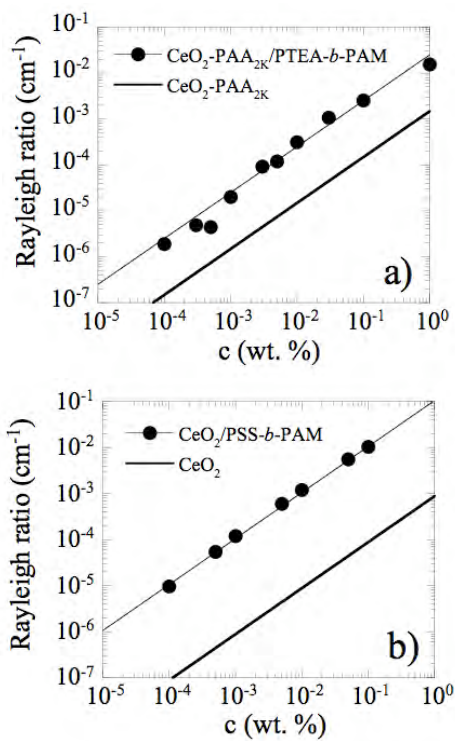

**Figure 3** : Same as in Figure 2 for a) CeO2-PAA2K/PTEA11K-b-PAM30K and b) CeO2/PSS7K-b-PAM30K formulated at c = 1 and c = 0.1 wt. % respectively.

Also shown in Figs. 2 are the scattering intensities calculated for the same solutions, assuming that surfactants and polymers were not associated (thick lines). Under such conditions, Eq. 4 extrapolated to zero wave-vector (q → 0) could be re-written :

$$\mathcal{R}_{UnAgg}(c) = K_{Surf}M_w^{Surf}c_{Surf} + K_{Pol}M_w^{Pol}c_{Pol} \qquad (6)$$

where   $c = c_{Surf} + c_{Pol}$. The definitions of the different parameters in Eq. 6 are the same than those given in the experimental section. For the three mixed systems investigated, DTAB/PANA$_{5K}$-*b*-PAM$_{30K}$, DTAB/PSS$_{7K}$-*b*-PAM$_{30K}$ and SDS/PTEA$_{11K}$-*b*-PAM$_{30K}$ the Rayleigh ratios calculated from Eq. 6 increased linearly with slopes $1.1×10^{-2}$, $3.1×10^{-2}$ et $0.93×10^{-2}$ cm$^{-1}$, i.e. around 100 times lower than those of the electrostatic complexes (cf. Table II). The good overall agreement between the scattering intensities in the low concentration range and the calculated $\mathcal{R}_{UnAgg}(c)$ strongly suggests that the colloidal complexes disassembled below a critical association concentration.

| | Coacervate complexes | $\mathcal{R}(c)/c^{(a)}$ cm$^{-1}$ | $\mathcal{R}(c)/c^{(b)}$ cm$^{-1}$ | M$_{w,app}$ g mol$^{-1}$ | D$_H$ nm | cac wt. % |
|---|---|---|---|---|---|---|
| organic | DTAB/PANA$_{5K}$-*b*-PAM$_{30K}$ | 1.3 | $1.1 \cdot 10^{-2}$ | $2.0×10^6$ | 60 | $1.1×10^{-2}$ |
| | DTAB/PSS$_{7K}$-*b*-PAM$_{30K}$ | 2.4 | $3.1 \cdot 10^{-2}$ | $5.9×10^6$ | 120 | $4.2×10^{-2}$ |
| | SDS/PTEA$_{11K}$-*b*-PAM$_{30K}^{(c)}$ | 0.95 | $0.93 \cdot 10^{-2}$ | $2.3×10^6$ | 75 | $5.0×10^{-3}$ |
| hybrid | CeO2-PAA$_{2K}$/PTEA$_{11K}$-*b*-PAM$_{30K}$ | 2.5 | $14.6 \cdot 10^{-2}$ | $5.5×10^6$ | 100 | $< 10^{-4}$ |
| | CeO2/PSS$_{7K}$-*b*-PAM$_{30K}$ | 10.6 | $11.4 \cdot 10^{-2}$ | $31×10^6$ | 150 | $< 10^{-4}$ |

**Table II** : Rayleigh ratios normalized by the concentration R(c)/c, molecular weight MW, hydrodynamic diameter DH and critical aggregation concentration cac for organic and hybrid coacervate complexes studied in this work. (a) : The data in the first column indicate the Rayleigh ratio that is measured in the high concentration range and arising from the aggregates. (b) : The second column estimates the scattering intensities if the nanocolloids and polymers would remain unassociated (Eqs. 6 and 7). These two regimes are illustrated by straight lines in Figs. 2 and 3. (c) : solution with no added salt.

### III.1.2 - Critical Association Concentration of hybrid Complexes

Figs. 3 show the concentration dependences of the Rayleigh ratio for CeO2-PAA2K/PTEA$_{11K}$-*b*-PAM30K and CeO2/PSS$_{7K}$-*b*-PAM$_{30K}$ formulated respectively at c = 1 and c = 0.1 wt. %. For both systems, a linear variation was found down to a concentration of $10^{-4}$ wt. %. Contrary to the data of the organic complexes, Figs. 3 did not exhibit a drop of the intensity upon dilution. For these two samples, dynamic light scattering has revealed a slightly polydisperse diffusive relaxation mode associated with hydrodynamic diameters D$_H$ = 100 ± 10 nm and 150 ± 10 nm, respectively. Cryo-TEM was also performed on CeO2- PAA$_{2K}$/PTEA$_{11K}$-*b*-PAM$_{30K}$ dilute sample, and an illustration of the hybrid aggregates is provided in Fig. 1c (lower panel). The photograph covers spatial field that is $0.36×0.50$ μm$^2$ and displays clusters of nanoparticles [13]. A large visual field was shown here in order to stress that the clusters were dispersed in solutions, a result which was consistent with the visual observations of the solutions. A closer inspection revealed that the aggregates were slightly anisotropic, with a fixed diameter around 20 nm and slight polydispersity in length and morphology. Elongated and branched aggregates were observed too in Fig. 1c, with length comprised between 20 and 100 nm. A statistical description of the clusters will be given in a forthcoming paper. In Figs. 3, the scattering intensities obtained from unassociated nanoparticles and polymers were shown for comparison (thick lines). These lines were obtained from Eq. 4 according to :

$$\mathcal{R}_{UnAgg}(c) = K_{Pol}M_w^{Pol}c_{Pol} + K_{Part}M_w^{Part}c_{Part} \qquad (7)$$

Here, the relevant contribution was that of the particles, which in both figures lied well below the experimental data. Their slopes were at $14.6 \times 10^{-2}$ and $11.4 \times 10^{-2}$ cm$^{-1}$ respectively (Table II). The observation of a linear dependence down to the lowest concentration available supports the assumption that the hybrids are more stable toward dilution than their organic counterparts.

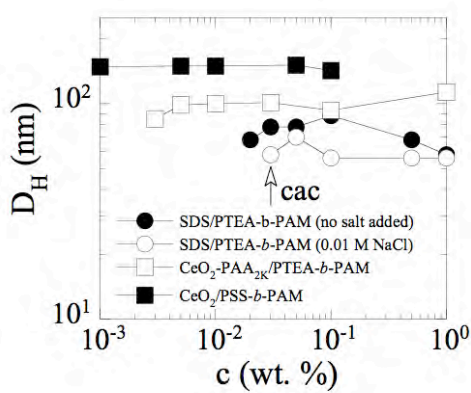

**Figure 4** : Hydrodynamic diameter DH as a function of the concentration for organic (SDS/PTEA11K-b-PAM30K, circles) and hybrid (CeO2-PAA2K/PTEA11K-b-PAM30K, open squares and CeO2/PSS7K-b-PAM30K, close squares) complexes. For the surfactant systems, the diameters were not determined below the cac (arrow).

Fig. 4 shows the evolution of the hydrodynamic diameter as a function of the concentration for SDS/PTEA$_{11K}$-b-PAM$_{30K}$, CeO$_2$-PAA$_{2K}$/PTEA$_{11K}$-b-PAM$_{30K}$ and CeO$_2$/PSS$_{7K}$-b-PAM$_{30K}$. For the surfactant based complexes, D$_H$ remained constant at 70 ± 10 nm down to the critical aggregation concentration. Below, the scattering intensity was too weak and its autocorrelation function could not be analyzed properly. For the nanoparticles based complexes, D$_H$ remained constant at 100 ± 10 nm and 150 ± 10 nm for CeO$_2$-PAA$_{2K}$/PTEA$_{11K}$-b-PAM$_{30K}$ and CeO$_2$/PSS$_{7K}$-b-PAM$_{30K}$, respectively. The main finding of the light scattering dilution studies was that in the range of observation, the organic complexes exhibited a cac whereas the hybrids did not.

### III. 2 – Adsorption kinetics

#### III.2.1 – Copolymers

Reflectometry experiments were performed at a concentration c = 0.1 wt. %. For the different systems investigated, this concentration insured that the adsorption occurred in the plateau region of the adsorption isotherm.50 Adsorption kinetics was first monitored on silica and poly(styrene) model substrates using PTEA$_{11K}$-b-PAM$_{30K}$ and PSS$_{7K}$-b-PAM$_{30K}$ polymeric solutions.

In dilute solutions, the copolymers are considered as dispersed and non-aggregated. On anionically charged silica surfaces, PTEA$_{11K}$-b-PAM$_{30K}$ were found to adsorb up to a value of Γ$_{ST}$ = 0.8 mg m$^{-2}$. This adsorption was most likely due to the electrostatic attractive interactions between the opposite charges coming from the cationic polyelectrolyte blocks and the surface. A subsequent rinsing with de-ionized water did not remove the copolymer layer. On the same substrate the anionic-neutral diblocks PSS$_{7K}$-b-PAM$_{30K}$ remained unadsorbed, at a stationary value Γ$_{ST}$ = 0 ± 0.1 mg m$^{-2}$.

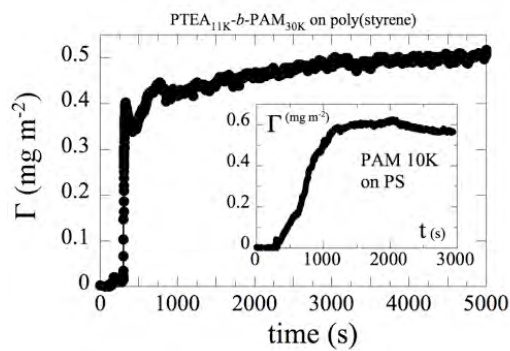

**Figure 5** : Adsorption kinetics for PTEA11K-b-PAM30K block copolymers on poly(styrene) substrate as received from stagnation point adsorption reflectometry. Measurements were performed with dilute solutions at c = 0.1 wt. %. Inset : adsorption kinetics for poly(acrylamide) molecular weight 10 000 g mol-1 operated in the same conditions. In both experiments, rinsing with de-ionized water did not remove the polymer layer.

In the case of a hydrophobic PS surfaces, adsorption was found for both polymers, at a lower level however than on silica surfaces (Γ$_{ST}$ = 0.4 - 0.6 mg m$^{-2}$). An example of adsorption kinetics is illustrated in Fig. 5 for PTEA$_{11K}$-b-PAM$_{30K}$. This is a surprising result since in the case a strong polyelectrolyte on an uncharged surface, electrostatics is not expected to contribute to the adsorption energy. Only at high salt content, the segment adsorption energy might compete with the segmental repulsion, leading eventually to adsorption [50]. More recently, cationic polyelectrolytes such as poly(vinylamine) were also found to noticeably adsorb on poly(styrene) substrates.38 In order to understand this phenomenon, reflectometry was also carried out in the same conditions on the different homopolymers with comparable molecular weights, namely PSS, PTEA and PAM. On hydrophobic PS surfaces, poly(acrylamide) with molecular weight 10000 g mol$^{-1}$ was the only polymer to adsorb spontaneously at a low but significant level (Γ$_{ST}$ = 0.5 mg m$^{-2}$, inset in Fig. 5). Again, in this particular case no desorption occurred on rinsing. These findings suggest that the copolymer adsorption shown in Fig. 5 occurs on PS through the PAM neutral chains. One possible explanation is the release of water molecules trapped in a frozen-like structure close to the hydrophobic surface. The system thus gains entropy by minimizing the contact area between the PS surface and the aqueous solution [21].

#### III.2.2 - Organic Coacervates

Figs. 6a and 6b show the adsorption features of the organic core-shell SDS/PTEA$_{11K}$-b-PAM$_{30K}$ system on anionically charged silica and hydrophobic poly(styrene) substrates, respectively. For both surfaces, adsorption occurred at a level of 1.2 mg m$^{-2}$ at steady state. Such values for the adsorbed amount are consistent with results obtained on electrostatic coacervate phases.36$^{-3}$8 As seen previously for the polymers alone, the causes of the adsorption might be different, depending on the substrate. Since PAM chains have no direct affinity to silica surfaces, the driven force for adsorption could be due either to the presence of residual positive charges associated to the SDS/PTEA$_{11K}$ cores, or possibly to the unfolding and reorganization of the core-shell structure on the local scale [21].

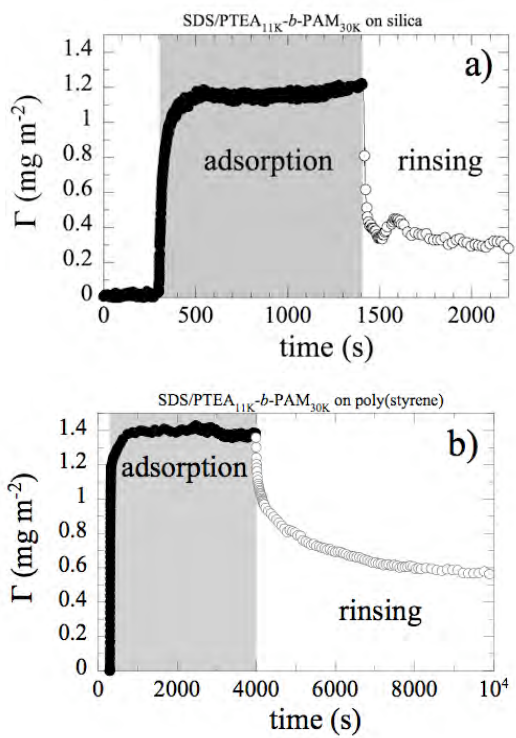

**Figure 6** : Adsorption kinetics for organic core-shell SDS/PTEA11K-b-PAM30K coacervates complexes on a) anionically charged silica and b) hydrophobic poly(styrene) substrate. Upon rinsing, the reflectometry signal dropped down to a level comprised between $0.3 - 0.5$ m-2. The decrease of the reflectometry signal suggests that the adsorbed layer has been partially removed during this second phase.

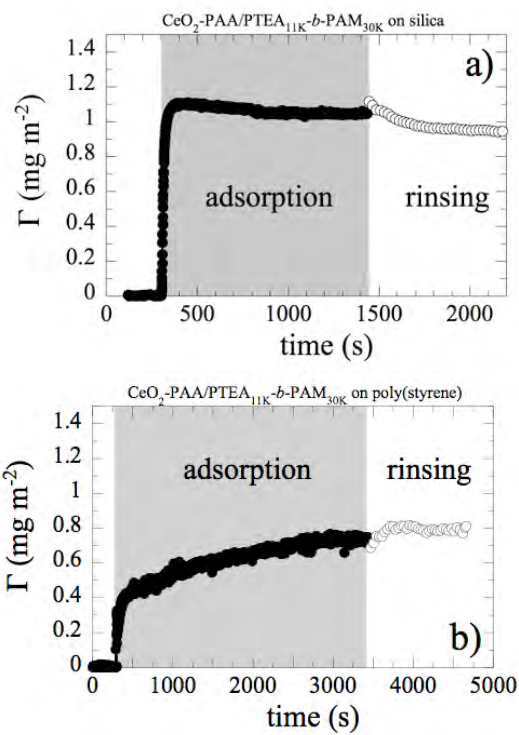

**Figure 7** : Same as in Fig. 6 for the CeO2-PAA2K/PTEA11K-b-PAM30K polymer-nanoparticle hybrids on silica (a) and on poly(styrene) (b) substrates. In contradistinction with the previous data, the deposited layer did not desorb upon rinsing.

In the case of PS on the contrary, the adsorption of coacervates is assumed to originate from the PAM-PS interactions disclosed in the case of the homo- and block copolymers. Upon rinsing (the second sequence of the kinetics), the reflectometry signal dropped significantly for both surfaces, although in the case of PS the final amount lied higher that for silica. These results suggest that the adsorbed layer has been partially removed during the rinsing stage.31 Since a finite amount remained on both surfaces (at a level comprised between $0.3 - 0.5$ mg m$^{-2}$), it was supposed that the original layer has undergone structural changes upon rinsing. With the experimental device used here, rinsing acts primarily as a dilution process. This in turn leads to a disassembling of the complexes, as seen in the bulk properties below the cac.

During the rinsing process, several scenarios can occur in addition to the disassembling of the complexes. For silica surfaces, it is plausible that the cationic PTEA blocks adsorb onto the oppositely charged substrate, leading to a finite adsorbed amount even at long times. In the case of PS, both SDS surfactants and PTEA$_{11k}$-b-PAM$_{30K}$ may remain simultaneously [58].

### III.2.3 - Hybrid Coacervates

Figs. 7 and 8 show the adsorption properties of the hybrid coacervates CeO$_2$-PAA$_{2K}$/PTEA$_{11K}$-b-PAM$_{30K}$ and CeO$_2$/PSS$_{7K}$-b-PAM$_{30K}$ on silica and on PS substrates. The three sets of data exhibit the same behavior : after an adsorption process at a level comparable to those found for the organic complexes (0.8 – 1.4 mg m$^{-2}$), no desorption occurred upon rinsing [31]. We explain the absence of desorption observed here as a consequence of the strong stability of the polymer-particle hybrids upon dilution, a result that was already emphasized in the previous section. Clearly, the existence of a very low cac for the particle-polymer hybrids (below 10$^{-4}$ wt. %, see Table II) guarantees the integrity of the deposited layer on the two substrates. The difference in the steady state adsorbed amounts between the two systems ($\Gamma$ST = 0.8 mg m$^{-2}$ for CeO$_2$-PAA$_{2K}$/PTEA$_{11K}$-b-PAM$_{30K}$ and $\Gamma$ST = 1.4 mg m$^{-2}$ for CeO$_2$/PSS$_{7K}$-b-PAM$_{30K}$) may arise from the different morphologies of the aggregates built up during complexation. Although weaker, the adsorbed amount on PS of CeO₂-PAA$_{2K}$/PTEA$_{11K}$-*b*-PAM$_{30K}$ is not negligible. After dipping a PS substrate into a 0.1 wt. % solution of those hybrids, the substrate became hydrophilic, as evidenced by the formation of a wetting film once the film was removed from the solution and further by a measure of the contact angle at θr < 30° (receding conditions).

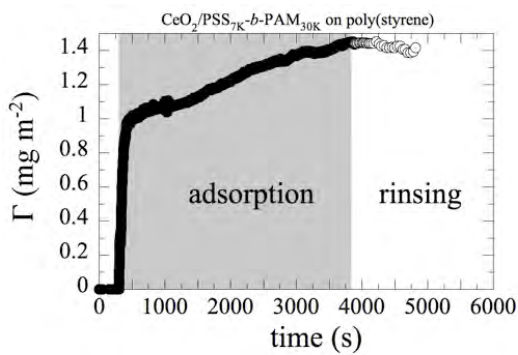

**Figure 8** : *Adsorption kinetics for CeO2/PSS7K-b-PAM30K hybrids on hydrophobic poly(styrene) substrate. Upon rinsing, the reflectometry signal remained as the same adsorbed level.*

## V – Conclusion

We have investigated the bulk and surface properties of mixed aggregates resulting from the co-assembly between nanocolloids (surfactant micelles, cerium oxide nanoparticles) and block copolymers. The association was driven by the opposite electrostatic charges borne by each constituent of the co-assembly. The use of copolymers allowed to confine the association at a size that remained in the colloidal domain, that is ~ 100 nm. Three systems made with surfactants, DTAB/PANa$_{5K}$-*b*-PAM$_{30K}$, DTAB/PSS$_{7K}$-*b*-PAM$_{30K}$ and SDS/PTEA$_{11K}$-*b*-PAM$_{30K}$, and two made from the nanoceria, CeO₂-PAA$_{2K}$/PTEA$_{11K}$-*b*-PAM$_{30K}$ and CeO₂/PSS$_{7K}$-*b*-PAM$_{30K}$, were investigated in order to demonstrate the generality of the results. Light scattering studies have shown that the complexes made with surfactants all exhibited a critical association concentration around 10⁻² wt. %. For one system, DTAB/PANa$_{5K}$-*b*-PAM$_{30K}$, the cac was corroborated by potentiometric measurements utilizing electrodes selective for the surfactant. For CeO₂-based complexes, no cac could be evidenced down to 10⁻⁴ wt. %. Cryo-TEM observations have disclosed unambiguously the cluster-like microstructure of the hybrid aggregates. For one system, CeO₂-PAA$_{2K}$/PTEA$_{11K}$-*b*-PAM$_{30K}$, the aggregates were shown to be slightly anisotropic with a lateral dimensions around 20 nm and total lengths comprised between 20 and 50 nm. These results compare very well to those found recently on oxide nanoparticles coated with citrates counterions [13].

The SPAR measurements have shown that organic and hybrid aggregates adsorbed readily on model substrates (as illustrated in Fig. 9). At steady state, the adsorbed amount attained values of the order of 1 – 1.5 mg m², which correspond to a single and densely packed monolayer of colloids. These estimations were made considering the extension of the core-shell aggregates as being delimited by their hydrodynamic diameters (Table II).

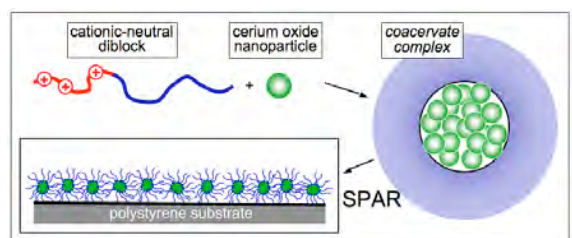

**Figure 9** : *Schematic representation of the formation of hybrid coacervate complexes and of their adsorption on model substrates. SPAR is the abbreviation for Stagnation Point Adsorption Reflectometry.*

In order to explain the unexpected adsorption of the hydrophilic aggregates on the hydrophobic poly(styrene), SPAR experiments were performed with the copolymers alone, as well as with the homopolymers from which these copolymers were composed. From these tests, it was recognized that poly(acrylamide) with molecular weight 10 K adsorbed spontaneously on PS, a result that was not reported for this substrate. This latter findings was important since it allowed us to conclude that the adsorption of the coacervates was primarily driven by that of the PAM corona. For the surfactant-based complexes and only for these systems, the pre-adsorbed layer was easily removed upon rinsing. The addition of water at the stagnation point acted primarily as a dilution process of the deposited layer. As seen for the bulk properties, for the surfactant system, dilution down to below the cac leads to a disassembling of the complexes, and therefore to its almost complete desorption. In conclusion, we have shown that a bulk property such as the critical aggregation concentration of coacervates translates into a useful surface property when dealing with hybrid structures. Beyond the sole durability of the coating offered by the hybrid structures, the presence of inorganic nanoparticles carrying their intrinsic properties will certainly permit the elaboration of coatings with multiple all-in-one functionalities.

**Acknowledgements** : The authors thank Rhodia for technical and financial support. We are indebted to Annie Vacher and Marc Airiau (Centre de Recherches d'Aubervilliers, Rhodia, France) for the cryo-TEM experiments on the nanoparticles and hybrids. Aurélie Bertin et Pascal Hervé are also acknowledged for the binding isotherms and measurements of the cacs.